\documentclass[letterpaper,twocolumn,10pt]{article}
\usepackage{usenix,epsfig,endnotes}
\usepackage[english]{babel}
\usepackage{blindtext}
\usepackage{graphicx,subcaption}
\usepackage{amsmath}
\usepackage{amsfonts}
\usepackage{amssymb}
\usepackage{xspace}
\usepackage{comment}
\usepackage{booktabs}
\usepackage{enumitem}
\usepackage{microtype}
\usepackage{multirow}
\usepackage[title]{appendix}
\graphicspath{{figs/}}

\newcommand{\sysname}{\textsc{DeDe}\xspace}
\newcommand{\sysnamestar}{\textsc{DeDe$^*$}\xspace}
\DeclareMathOperator*{\argmin}{arg\,min}

\DeclareMathOperator*{\maximize}{Maximize~}
\DeclareMathOperator*{\minimize}{Minimize~}
\newcommand{\parab}[1]{\smallskip\noindent\textbf{#1}~}
\newcommand{\parai}[1]{\smallskip\noindent\textit{#1}~}

\usepackage{listings}

\usepackage[colorinlistoftodos,prependcaption,textsize=tiny]{todonotes}

\usepackage{pifont}
\newcommand{\cmark}{\ding{51}}

\definecolor{codegreen}{rgb}{0,0.6,0}
\definecolor{codeblue}{rgb}{0,0,0.8}
\definecolor{codegray}{rgb}{0.5,0.5,0.5}
\definecolor{codepurple}{rgb}{0.58,0,0.82}
\definecolor{codered}{rgb}{0.8,0,0}
\definecolor{codebrown}{rgb}{0.75,0.5,0.1}

\lstdefinelanguage{myLang}{
    keywords=[1]{import,as,for,in},
    keywordstyle=[1]\color{codepurple},
    keywords=[2]{Variable,Parameter,Maximize,Problem,solve,sum},
    keywordstyle=[2]\color{codeblue},
    morecomment=[l]{\#},
    commentstyle=\color{codegreen},
    stringstyle=\color{red}
}
\lstdefinestyle{myStyle}{
    backgroundcolor=\color{white},
    basicstyle=\fontfamily{lmtt}\small,
    basewidth=0.5em,
    breakatwhitespace=false,
    numberstyle=\scriptsize\color{codegray},
    framerule=0.5pt,
    breaklines=true,
    keepspaces=true,
    numbers=left,
    xleftmargin=1.5em,
    frame=single,
    framexleftmargin=1.5em,
    numbersep=5pt,
    showspaces=false,
    showstringspaces=false,
    showtabs=false,
    tabsize=2,
    language=myLang,
}

\newcommand{\autha}{\rm\textsuperscript{\textdagger}}
\newcommand{\authb}{\rm$^{\mathbb{I}}$}

\begin{document}

\newcommand{\titlespace}{\hspace{10pt}}
\title{\Large \bf Decouple and Decompose: Scaling Resource Allocation with \sysname\\[-5pt]}

\author{
\rm Zhiying Xu\autha \titlespace Minlan Yu\autha \titlespace Francis Y. Yan\authb\\
{\fontsize{11pt}{13pt}\selectfont \autha\textit{Harvard University}} \titlespace {\fontsize{11pt}{13pt}\selectfont \authb\textit{University of Illinois Urbana-Champaign}}
}
\date{}

\maketitle

\pagestyle{empty}
\thispagestyle{empty}

{
\let\origthefootnote\thefootnote
\let\thefootnote\relax
\footnote{This work is accepted at USENIX OSDI 2025.
}
\let\thefootnote\origthefootnote
}
\vspace{-0.2in}
\begin{abstract}
Efficient resource allocation is essential in cloud systems to
facilitate resource sharing among tenants.
However, the growing scale of these optimization problems
have outpaced commercial solvers commonly employed in production.
To accelerate resource allocation,
prior approaches either customize solutions for narrow domains
or impose workload-specific assumptions.
In this work, we revisit real-world resource allocation problems
and uncover a common underlying structure:
the vast majority of these problems are inherently \textit{separable},
i.e., they optimize the aggregate utility of individual
resource and demand allocations, under separate constraints for
each resource and each demand.
Building on this observation, we develop \sysname, a scalable and
theoretically rooted optimization framework for large-scale resource allocation.
At the core of \sysname is a \textit{decouple-and-decompose} approach:
it decouples entangled resource and demand constraints
and thereby decomposes the overall optimization into alternating
per-resource and per-demand subproblems
that can be solved efficiently and in parallel.
We have implemented and released \sysname as a Python package
with a familiar modeling interface.
Our experiments on three representative resource allocation
tasks---cluster scheduling, traffic engineering, and load
balancing---demonstrate that \sysname delivers significant speedups
while generating higher-quality allocations.
\end{abstract}

\section{Introduction}
\label{sec:intro}

Resource allocation remains a critical and challenging problem today,
especially as cloud providers operate multi-tenant systems on an
unprecedented scale---these systems must ensure efficient and fair allocation of
computing, storage, and networking resources across a large number of clients,
in order to maintain the quality of cloud services.

Common scenarios of resource allocation include
cluster scheduling~\cite{newell2021ras,jayaram2023sia,narayanan2020heterogeneity, xiao2018gandiva,yang2023skypilot},
traffic engineering~\cite{hong2013achieving,abuzaid2021contracting,zhong2021arrow,singh2021cost2,krishnaswamy2022decentralized},
and load balancing~\cite{curino2011workload, serafini2014accordion, taft2014store, lee2021shard}.
To meet diverse and dynamic user demands
(jobs, traffic flows, and queries),
these systems extensively employ commercial optimization solvers
like Gurobi~\cite{gurobi}
to solve linear programs (LP) or mixed-integer linear programs (MILP)
that determine the allocation of valuable cloud resources
(compute units, network links, and data nodes)
worth millions to billions of dollars.

However, as cloud environments continue to expand and diversify,
the sheer scale 
of modern resource allocation problems has exceeded the capabilities of commercial solvers.
These problems may involve millions of variables
and take tens of minutes or even hours to solve,
whereas fast allocation is necessary to meet service-level objectives.
This ``scalability crisis'' has spurred
a flurry of studies~\cite{abuzaid2021contracting,xu2023teal,perry2023dote,namyar2023solving}
that trade off solution quality for computational speed,
using heuristics, approximate algorithms, or machine learning.
Nevertheless, these approaches are typically restricted to specific
domains (e.g., wide-area network traffic engineering)
or particular optimization objectives (e.g., max-min fairness),
limiting their broader applicability.
While POP~\cite{narayanan2021solving} 
accelerates assorted allocation tasks,
it hinges on a
``granular'' assumption that each demand requests only a small fraction of
interchangeable resources.
This premise proves brittle under realistic workloads, leading to degraded
solution quality (\S\ref{sec:eval}).

\begin{figure}[t]
    \centering
    \includegraphics[width=\columnwidth]{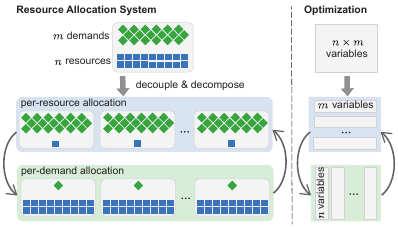}
    \vspace{-20pt}
    \caption{Overview of \sysname.}
    \label{fig:dede_overview}
    \vspace{-10pt}
\end{figure}

In this study, we revisit real-world resource allocation problems
through a different lens, avoiding domain-specific designs or
workload-dependent assumptions.
We introduce \sysname\footnotemark[1]\footnotetext[1]{\sysname encapsulates
our core technique ``\textbf{De}couple and \textbf{De}compose''.},
a practical and scalable optimization framework for solving
resource allocation problems.
Rooted in optimization theory,
\sysname allows the global allocation problem to be decomposed
into alternating per-resource and per-demand subproblems
(Figure~\ref{fig:dede_overview}),
which can be solved more efficiently and in parallel
without compromising solution quality.

\begin{figure*}[t]
    \centering
    \includegraphics[width=0.9\textwidth]{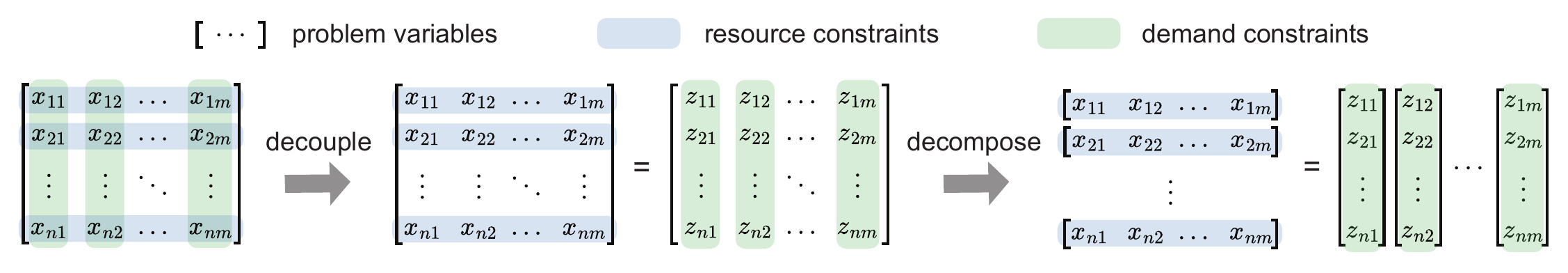}
    \vspace{-10pt}
    \caption{\sysname decouples constraints using an auxiliary variable,
    thereby decomposing optimization into subproblems.}
    \label{fig:dede_workflow}
    \vspace{-10pt}
\end{figure*}

This decomposition is enabled by an intrinsic, workload-agnostic
problem structure that we identified after surveying dozens of
real-world resource allocation problems
in the literature (Table~\ref{tab:comparison}).
We find that the vast majority of these problems
are inherently \textit{separable}:
they optimize the total utility of individual resource and demand allocations,
subject to separate constraints for each resource and demand
to ensure feasibility.
To be more concrete, we illustrate this problem structure using
a simplified job scheduling example.

\parab{$\blacktriangleright$ Example:}
Each hour, we aim to determine---as soon as possible---the optimal
allocation of a set of large language model
inference jobs (e.g., GPT-4, Llama 3, and DeepSeek-V3)
on a heterogeneous GPU cluster composed of various hardware types
(e.g., Nvidia H100, A100, and V100).
Jobs may divide their execution time across multiple GPU types,
and due to differences in hardware specifications (e.g., memory capacity, bandwidth,
and FLOPS), each job may require a different number of GPUs for each
GPU type and achieve varying levels of throughput,
measured in tokens per second (TPS).
For instance, a job might run for 0.4 hours on 2$\times$A100 at 50 TPS,
and for 0.6 hours on 4$\times$V100 at 20 TPS.

Formally, let $x_{ij}$ denote the fraction of the hour that job $j$ (demand)
runs on GPU type $i$ (resource).
Naturally, we have the demand constraints:
$\forall j, \sum_i x_{ij} \le 1$.
On the other hand, for each GPU type, the total GPU hours requested by all jobs
must not exceed the available GPU capacity.
Thus, the resource constraints are:
$\forall i, \sum_j \text{req}_j \cdot x_{ij} \le \text{capacity}_i$.
The objective is to maximize the total weighted average
throughput, $\sum_j w_j \cdot (\sum_i \text{tput}_{ij} \cdot x_{ij})$,
reflecting different job priorities. $\square$

\smallskip
This example demonstrates the ``separable'' structure
frequently encountered in practical resource allocation problems:
resource constraints are specified for \textit{each} resource and its
associated demands, demand constraints are specified for
\textit{each} demand and its contributing resources,
and the objective function sums the utility achieved by
\textit{each} demand (or resource).
Despite this seemingly modular structure, however,
such optimization problems remain
computationally challenging for commercial solvers to solve at scale.
The difficulty arises from each allocation variable $x_{ij}$
simultaneously affecting both resource and demand constraints,
precluding straightforward decomposition and parallel optimization.

To address this challenge, \sysname first \textit{decouples}
entangled resource and demand constraints through a mathematical
reformulation. It introduces a copy of the allocation matrix $x$ as an
auxiliary variable $z$,
and rewrites the demand constraints (and per-demand terms in the objective)
in terms of $z$, while adding a new constraint $x=z$.
This transformation is mathematically equivalent, preserving
the optimal solution.
Moreover, it also lends itself to ADMM (Alternating Direction Method
of Multipliers)~\cite{admm}, a constrained optimization method based
on Lagrange multipliers~\cite{bertsekas2014constrained}.
By incorporating constraints into the objective via multipliers
and constructing the (augmented) Lagrangian (see \S\ref{sec:decouple}),
ADMM provides a theoretical framework that allows \sysname to alternatively
and iteratively optimize the Lagrangian with respect to each block of
variables---$x$ or $z$---while holding the other fixed.

Building on the reformulation, \sysname further \textit{decomposes} the
optimization into subproblems along each resource and each demand.
This step capitalizes on the identified separable structure,
wherein the introduced Lagrangian can be
partitioned across individual resources and demands (see \S\ref{sec:decompose}).
Consequently, instead of solving a monolithic optimization over
all $n$ resources and $m$ demands,
\sysname breaks the problem into $n$ per-resource subproblems,
each with only $m$ demand variables,
and $m$ per-demand subproblems, each with only $n$ resource variables.
These smaller subproblems
are amenable to independent and parallel optimization using off-the-shelf
solvers.
Figure~\ref{fig:dede_workflow} depicts the 
decouple-and-decompose workflow of \sysname.

To demonstrate \sysname's practical utility,
we have implemented \sysname as a Python package,
installable via \texttt{pip install dede}.
Built on the popular open-source modeling language
\texttt{cvxpy}~\cite{cvxpy1, cvxpy2},
\sysname provides familiar APIs (\S\ref{sec:implementation})
aligned with \texttt{cvxpy}.
Moreover, unlike prior efforts that simulate
parallelism~\cite{abuzaid2021contracting, narayanan2021solving},
our implementation is optimized for \textit{real} parallel execution using 
Ray~\cite{moritz2018ray}, enabling efficient utilization of many CPU cores
with minimal overhead.

We evaluate \sysname on three representative resource allocation
tasks---cluster scheduling, traffic engineering, and load
balancing, demonstrating its faster speed and higher allocation
quality relative to the state of the art (\S\ref{sec:eval}).
Compared with the best variant of POP~\cite{narayanan2021solving}
in each domain,
\sysname achieves a 7.3\% improvement in allocation quality and
a 3.1$\times$ speedup for cluster scheduling,
5.3\% and 7.6$\times$ for traffic engineering,
and 12.6\% and 2.2$\times$ for load balancing.

\section{Real-World Resource Allocation Problems}
\label{sec:motivation}\label{sec:formulation}

In modern cloud systems, resources such as CPUs, memory,
and network bandwidth are shared among numerous users and applications.
A central resource allocator often casts the distribution of resources
as an optimization problem (e.g., LP or MILP) and
repeatedly computes feasible solutions using solvers
to accommodate changing demands.

However, these solvers are struggling to keep up with the
increasing problem sizes that frequently occur in fast-growing systems.
They may take up to \textit{several hours} to solve
an allocation problem that attempts to assign thousands
of resources to thousands of
demands. This is several orders of magnitude slower than desired
(e.g., service-level objectives measured in seconds).
This challenge of scaling up resource allocation has prompted
a series of recent works~\cite{abuzaid2021contracting,narayanan2021solving,
xu2023teal,perry2023dote,namyar2023solving},
but they either rely on domain-specific customizations or
workload-dependent assumptions.
In this work, we instead seek to take a
general (i.e., domain- and workload-agnostic) approach
to scaling resource allocation.

\parab{Separable problem structure.}
Upon surveying real-world resource allocation problems from
the literature (Table~\ref{tab:comparison}),
we find that nearly all these formulations can be
transformed into a \textit{separable} structure (\S\ref{sec:intro})
formally described as follows.
(We discuss the generality and limitations in \S\ref{sec:applicability}.)

\begin{itemize}[noitemsep,topsep=0pt,leftmargin=*]

\item
\parai{Variables.} 
An allocation matrix $x \in \mathcal{X}$ specifies the allocation of 
$n$ resources among $m$ demands, where $x_{ij}$ indicates the amount
or fraction of resource $i$ assigned to demand $j$ (or vice versa).
Accordingly, the $i$-th row of $x$, denoted as
$x_{i*} = (x_{i1}, x_{i2}, \dots, x_{im})$,
constitutes the allocation vector of resource $i$, while
the $j$-th column $x_{*j} = (x_{1j}, x_{2j}, \dots, x_{nj})$ encodes
the allocation of demand $j$.
The constraint set $\mathcal{X}$ is the Cartesian product of the individual
domains $\mathcal{X}_{ij}$, but it is often just non-negative real numbers
or integers.

\item
\parai{Objective.} The objective defines a metric to be maximized or minimized
over an allocation matrix $x$, consisting of a sum of
utilities or costs that quantify the outcome
from allocating each resource $x_{i*}$ or demand $x_{*j}$:
\begin{equation}
\sum_{i} f_i(x_{i*}) + \sum_{j} g_j(x_{*j}), \label{eq:obj}
\end{equation}
where $f_i(\cdot):\mathbb{R}^m \rightarrow \mathbb{R}$ and 
$g_j(\cdot):\mathbb{R}^n \rightarrow \mathbb{R}$ are utility or cost functions.
These functions are typically convex (by nature or by choice), facilitating
tractable optimization.
A variety of objectives can be transformed into this form using standard optimization
techniques. E.g., maximizing the minimum utility can be converted into maximizing
an auxiliary ``min utility'' variable, with additional constraints ensuring that each utility
is at least as large as this variable.

\item
\parai{Constraints.}
To enforce feasible allocation,
one or more constraints can be imposed on each resource and each
demand (e.g., to prevent oversubscription of resources
and overprovisioning of demands).
For computational tractability, these constraints are typically linear.
Following standard optimization practices,
we transform inequality constraints into equivalent equality constraints
using slack variables.
Thus, without loss of generality, we express the $rc_i\ge 1$ resource constraints
on each resource $i$ as
\begin{equation}
R_i x_{i*} = r_i, \quad \forall i, \label{eq:resource_constr}
\end{equation}
where $R_i\in \mathbb{R}^{rc_i \times m}, r_i\in \mathbb{R}^{rc_i}$ are constraint
parameters.
Likewise, each demand $j$ is subject to $dc_j\ge 1$ linear constraints:
\begin{equation}
D_j x_{*j} = d_j, \quad \forall j, \label{eq:demand_constr}
\end{equation}
where $D_j\in \mathbb{R}^{dc_j \times n}, d_j\in \mathbb{R}^{dc_j}$ are
constraint parameters.
\end{itemize}

Despite the separable objective and separate constraints over each resource and
each demand, large resource allocation problems (e.g., with millions of
variables) still remain painfully slow to solve as they cannot actually be
separated into smaller, concurrent subproblems.  The main obstacle is that
resource and demand constraints are fundamentally entangled.
As shown in Equations~\ref{eq:resource_constr} and~\ref{eq:demand_constr}, each
$x_{i,j}$ not only participates in the resource constraints (over $x_{i*}$) but
also in the demand constraints (over $x_{*j}$).

\begin{figure}[t]
    \centering
    \includegraphics[width=\columnwidth]{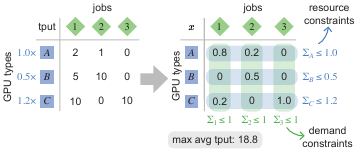}
    \vspace{-20pt}
    \caption{Toy scenario of the job scheduling example.}
    \label{fig:toy_example}
    \vspace{-10pt}
\end{figure}

\parab{$\blacktriangleright$ Example:}
The example introduced in \S\ref{sec:intro} clearly fits into
this formulation. For a more specific example, consider a toy scenario
in Figure~\ref{fig:toy_example}, where three language model inference jobs
are scheduled across three types of GPUs.
For simplicity, we assume each job fits in a single GPU
of any type ($\text{req}_j=1$), and all jobs have equal priority
($w_j=1$).

The left panel of Figure~\ref{fig:toy_example} shows the capacity
of each GPU type ($\text{capacity}_i$) in GPU hours, alongside the throughput table
($\text{tput}_{ij}$).
E.g., GPU type $C$ has a total capacity of 1.2 GPU hours, and assigning
job 1 to it yields a throughput of 10 TPS.

The right panel shows the optimal allocation matrix ($x_{ij}$),
which satisfies all resource and demand constraints.
E.g., the optimal solution schedules job 1 on
(one unit of) GPU type $A$ for 0.8 hours, and on GPU type $C$ for 0.2 hours.
The maximum average throughput achieved is 18.8 TPS. $\square$

\section{\sysname: Decouple and Decompose}
\label{sec:design}

We present \sysname, a practical and theoretically rooted optimization framework
for solving large-scale, real-world resource allocation problems.
By exploiting the inherently separable structure of these problems as described
earlier,
\sysname achieves massive parallelism while producing near-optimal solutions
through a \textit{decouple-and-decompose} approach.

\vspace{-5pt}
\subsection{Decouple}~\label{sec:decouple}
\vspace{-10pt}

The interdependence between resource and demand constraints is
a core complexity in resource allocation.
Achieving an optimal solution requires jointly allocating all resources to all
demands; otherwise, local, greedy strategies likely result in suboptimal or infeasible
solutions. 

Existing solutions do not address this intertwined structure of constraints in
resource allocation.
In practice, production systems tend to simply rely on commercial solvers and employ
linear programming (LP) or mixed-integer linear programming (MILP)
algorithms
(e.g., simplex methods~\cite{simplex}, barrier methods~\cite{barrier},
or branch-and-bound methods~\cite{boyd2007branch})
to navigate a high-dimensional search space, which is induced by
the coupling of resource and demand constraints.

A recent work, POP~\cite{narayanan2021solving}, naively breaks entangled constraints
by partitioning resources and demands into a handful of subsets (e.g., 4--64),
and then independently allocates a subset of resources to a subset of demands.
Doing so ignores the dependencies across subsets and leads to suboptimal
solutions in practice (\S\ref{sec:eval}).
Meanwhile, it only enables limited parallelism because finer-grained
decomposition would render the allocation infeasible and unusable.

Instead of disregarding the constraints across subproblems,
we tackle the root cause of constraint entanglement.
Specifically, without sacrificing solution quality,
our high-level goal is to decouple demand constraints from resource constraints,
transforming the global allocation into alternating
per-resource and per-demand subproblems.

\parab{Reformulation.}
Consider the minimization of the objective in Equation~\ref{eq:obj}
(the signs of $f_i$ and $g_j$ can be simply reversed for maximization problems),
subject to resource constraints in Equation~\ref{eq:resource_constr}
and demand constraints in Equation~\ref{eq:demand_constr}.

First, we introduce an auxiliary variable $z$
as a duplicate of the allocation matrix $x$,
and rewrite all demand constraints and all demand-related terms
in the objective using $z$. Meanwhile,
we add a new constraint $x=z$.
This transformation clearly leads to an equivalent optimization problem:
\begin{align}\label{eq:reformulated_opt}
\min_{x \in \mathcal{X}} 
\quad & \sum_{i} f_i( x_{i*} )+ 
\sum_{j} g_j( z_{*j}), \nonumber \\
s.t. \quad & R_i  x_{i*} = r_i, \quad \forall i, \\
& D_j  z_{*j} = d_j, \quad \forall j, \nonumber \\
& x - z = 0. \nonumber
\end{align}
In this formulation, all resource constraints remain on $x$,
while all demand constraints (and related utilities) are moved to $z$.

Although introducing $z$ appears to separate the problem
structurally, the (necessary) constraint $x=z$ still tightly couples the variables
and thus prevents straightforward decomposition.
In fact, jointly optimizing $x$ and $z$ 
(e.g., using penalty methods~\cite{penalty}
or augmented Lagrangian~\cite{augmentedlgr})
will forfeit the benefits of this reformulation,
as we will show in \S\ref{sec:eval_micro}.

\parab{ADMM background.}
Nevertheless, this reformulation lends itself to
ADMM (Alternating Direction Method of Multipliers)~\cite{admm},
a classical method for solving constrained optimization problems
of the form: minimize $f(x)+g(z)$, subject to $Ax+Bz=c$.
Note that Equation~\ref{eq:reformulated_opt} conforms to this form.

ADMM is based on augmented Lagrangian~\cite{augmentedlgr}, a method that
converts constrained optimization into a sequence of unconstrained
optimization problems using Lagrange
multipliers~\cite{bertsekas2014constrained} (a well-known technique in optimization).
These unconstrained problems are defined over the Lagrangian function,
constructed from the objective and constraints:
$L_\rho(x, z, y) = f(x) + g(z) + y^\mathsf{T}(Ax + Bz - c) + (\rho/2) \lVert Ax + Bz - c \rVert^2_2$,
where $y$ is (an estimate of) the Lagrange multiplier,
and the last quadratic term with parameter $\rho$ serves as a
penalty~\cite{penalty} for improving convexity and stability.

The theory of ADMM enables partial, alternating optimization of the Lagrangian
with respect to $x$ and $z$---iteratively solving for $x$ with $z$ fixed,
and solving for $z$ with $x$ fixed.
This ``two-block partial optimization''
guarantees convergence to an optimal solution for convex
problems~\cite{admm2blocks1, admm2blocks2}.
In non-convex settings (including those involving integers),
empirical results indicate that ADMM can still effectively solve
many of these problems~\cite{admmint, admmempirical, admmempirical2}.
Theoretical convergence results are established under certain
conditions~\cite{admmnonconvex, admmnonconvex2, admmnonconvex3}.

By introducing the auxiliary variable $u=(1/\rho)y$,
the augmented Lagrangian can be further rewritten as:
$L_\rho(x, z, u) = f(x) + g(z) + (\rho/2) \lVert Ax + Bz - c + u \rVert^2_2 - (\rho/2) \lVert u\rVert^2_2$, leading to a more convenient \textit{scaled form} of ADMM.
Next, we present the ADMM iterates for updating $x$, $y$, and $u$ directly
in the context of resource allocation problems.

\parab{Applying ADMM.}
\sysname applies the scaled form of ADMM to solve the reformulated optimization
problem (Equation~\ref{eq:reformulated_opt}).
First, it constructs the augmented Lagrangian:
\begin{equation}
\begin{split}
& \mathcal{L}_\rho(x, z, \alpha, \beta, \lambda) 
 = \sum_i f_i(x_{i*}) + \sum_j g_j (z_{*j}) \\
    &+ \frac{\rho}{2} \Bigl( \sum_i \lVert R_i x_{i*} - r_i + \alpha_i \rVert_2^2
                      + \sum_j \lVert D_j z_{*j} - d_j + \beta_j \rVert_2^2 \\
                      &+ \lVert x - z + \lambda \rVert_F^2 \Bigr) 
    - \frac{\rho}{2} (\lVert \alpha \rVert_2^2 + \lVert \beta \rVert_2^2 + \lVert \lambda \rVert_F^2),
\end{split}
\end{equation}
where $\alpha_i \in \mathbb{R}^{rc_i}$, $\beta_j \in \mathbb{R}^{dc_j}$,
$\lambda \in \mathbb{R}^{n\times m}$ are the introduced auxiliary variables.
$\lVert \cdot \rVert_2$ still denotes
the $\ell^2$-norm of a vector,
while $\lVert \cdot \rVert_F$ is
the Frobenius norm of a matrix, i.e., the $\ell^2$-norm of the matrix
when flattened into a vector.

The ADMM iterates are given by:
\begin{align}
x^{(k+1)} &:= \argmin_{x \in \mathcal{X}}\,\mathcal{L}_\rho(x, z^{(k)}, \alpha^{(k)}, \beta^{(k)}, \lambda^{(k)}) \label{eq:x_iter}\\
z^{(k+1)} &:= \argmin_z\,\mathcal{L}_\rho(x^{(k+1)}, z, \alpha^{(k)}, \beta^{(k)}, \lambda^{(k)}) \label{eq:y_iter} \\
\alpha_i^{(k+1)} &:= \alpha_i^{(k)} + R_i x^{(k+1)}_{i*} - r_i, \forall i \nonumber \\ 
\beta_j^{(k+1)} &:= \beta_j^{(k)} + D_j z^{(k+1)}_{*j} - d_j, \forall j \nonumber\\ 
\lambda^{(k+1)} &:= \lambda^{(k)} + x^{(k+1)} - z^{(k+1)} \nonumber
\end{align}
We will show in \S\ref{sec:decompose} that both
the optimization subproblems in Equations~\ref{eq:x_iter} and~\ref{eq:y_iter}
are amenable to decomposition owing to the previously identified separable problem structure.

\subsection{Decompose}~\label{sec:decompose}
\vspace{-10pt}

Up to this point, we have reformulated the original resource allocation problem into
alternating minimization steps of the augmented Lagrangian
with respect to $x$ and $z$.
Next, we take the $x$-minimization step as an example to describe
how the optimization is decomposed ($z$-minimization is similar).

\begin{table*}[t]
\centering
\begin{center}
{
\small
\aboverulesep=0ex
\belowrulesep=0ex
\renewcommand{\arraystretch}{1.25}
\begin{tabular}{p{100mm}|ccc|cc}
\toprule
& \multicolumn{3}{c|}{Variables} & \multicolumn{2}{c}{Objective}\\\cline{2-6}
& Boolean & Integer & Float & Linear & Convex \\\midrule
RDC\cite{wang2022rdc} & \cmark & & &\cmark &\\\midrule
SkyPilot\cite{yang2023skypilot} & \cmark & & & & \cmark\\\midrule
ARROW\cite{zhong2021arrow}, FlexWAN~\cite{miao2023flexwan} & \cmark & \cmark & &\cmark &\\\midrule
Shoofly\cite{singh2021cost} & & \cmark & \cmark &\cmark & \\\midrule
PODP\cite{baruah2022parallelism}, RAS\cite{newell2021ras}, Skyplane\cite{jain2023skyplane}, Oort\cite{lai2021oort}, TACCL\cite{shah2023taccl}, Shard Manager\cite{lee2021shard}, Zeta\cite{zhang2022zeta}, CASCARA\cite{singh2021cost2}, Sia\cite{jayaram2023sia}, POP\cite{narayanan2021solving} & \multirow{2}{*}{\cmark} & & \multirow{2}{*}{\cmark} & \multirow{2}{*}{\cmark} & \\\midrule
NetHint\cite{chen2022nethint}, Gavel\cite{narayanan2020heterogeneity}, Teal\cite{xu2023teal}, ONEWAN\cite{krishnaswamy2023onewan}, BLASTSHIELD\cite{krishnaswamy2022decentralized}, NCFlow\cite{abuzaid2021contracting}, Cerebro\cite{zheng2021cerebro}, DOTE\cite{perry2023dote},
POP\cite{narayanan2021solving} & & & \multirow{2}{*}{\cmark} & \multirow{2}{*}{\cmark} & \\\midrule
PCF\cite{jiang2020pcf}, Electricity Pricing\cite{werner2021pricing}, POP\cite{narayanan2021solving} & & & \cmark & & \cmark \\
\bottomrule
\end{tabular}}
\end{center}
\vspace{-10pt}
\caption{Real-world resource allocation problems in recent literature.} \label{tab:comparison}

\vspace{-5pt}
\end{table*}

Because the $x$-minimization step solves for $x$ only,
we can ignore all terms in the augmented Lagrangian without $x$
(such as demand constraints).
Thus, we rewrite Equation~\ref{eq:x_iter} as
\begin{align}
x^{(k+1)} =& \argmin_{x \in \mathcal{X}}\,\mathcal{L}_\rho(x, z^{(k)}, \alpha^{(k)}, \beta^{(k)}, \lambda^{(k)}) \tag{\ref{eq:x_iter}}\\
          =& \argmin_{x \in \mathcal{X}}\,\sum_i f_i(x_{i*}) + \frac{\rho}{2} \sum_i \big\| R_i x_{i*} - r_i + \alpha_i^{(k)} \big\|_2^2 \nonumber
             \\&+ \frac{\rho}{2} \big\| x - z^{(k)} + \lambda^{(k)} \big\|_F^2 \nonumber\\
          =& \argmin_{x \in \mathcal{X}}\,\sum_i \Bigl( f_i(x_{i*}) + \frac{\rho}{2} \big\| R_i x_{i*} - r_i + \alpha_i^{(k)} \big\|_2^2 \nonumber
             \\&+ \frac{\rho}{2} \big\| x_{i*} - z_{i*}^{(k)} + \lambda_{i*}^{(k)} \big\|_2^2 \Bigr) \nonumber.
\end{align}
Given that the new objective is a sum of disjoint terms across index $i$,
its minimization can be decomposed into independent, per-resource subproblems
over each $x_{i*}$:
\begin{equation}\label{eq:resource_subproblem}
\begin{split}
x_{i*}^{(k+1)} =& \argmin_{x_{i*}\,\in \mathcal{X}_{i*}}\,\Bigl(f_i(x_{i*}) + \frac{\rho}{2} \big\| R_i x_{i*} - r_i + \alpha_i^{(k)} \big\|_2^2
                 \\&+ \frac{\rho}{2} \big\| x_{i*} - z_{i*}^{(k)} + \lambda_{i*}^{(k)} \big\|_2^2 \Bigr), \quad \forall i.
\end{split}
\end{equation}
This decomposition is made possible by the separable structure:
(1) the original objective is separable, with a utility $f_i(x_{i*})$ defined
for allocating each resource $i$,
and (2) the resource constraints are likewise separable, as each constraint
$R_i x_{i*} = r_i$ pertains only to the corresponding resource $i$.

As a result, the original optimization problem over $n\times m$ variables
is broken down into $n$ independent subproblems, each involving only $m$ variables.
\sysname solves each subproblem using off-the-shelf solvers, which return either
exact or approximate solutions depending on the nature of $f$ (convex or not),
whether $\mathcal{X}$ contains integers, etc.
Since this decomposition allows up to $n$ parallel computations---often
numbering in the thousands in large-scale settings---\sysname is capable of
delivering substantial speedups assuming sufficient CPUs.

For linear programs, the original problem has a worst-case time
complexity of approximately
$O((n\cdot m)^{2.373})$~\cite{cohen2021solving,narayanan2021solving}.
After decomposition, \sysname processes $n$ subproblems, each with
a complexity of $O(m^{2.373})$, reducing the overall complexity to
$O(n\cdot m^{2.373})$. Nonetheless, this theoretical improvement assumes
that ADMM converges within
a constant number of iterations. In practice, the number of iterations
depends on the choice of the penalty parameter $\rho$ (which acts like a ``learning rate'') and other factors that influence convergence.

Similarly, the $z$-minimization step decomposes into independent, per-demand allocation
subproblems over each $z_{*j}$:
\begin{equation}\label{eq:demand_subproblem}
\begin{split}
z_{*j}^{(k+1)} =& \argmin_{z_{*j}}\,\Bigl( g_j(z_{*j}) + \frac{\rho}{2} \big\| D_j z_{*j} - d_j + \beta_j^{(k)} \big\|_2^2
                 \\&+ \frac{\rho}{2} \big\| x^{(k+1)}_{*j} - z_{*j} + \lambda_{*j}^{(k)} \big\|_2^2 \Bigr), \quad \forall j.
\end{split}
\end{equation}

\section{Generality and Limitations}
\label{sec:applicability}

In this section, we discuss the generality of
\sysname (conditions under which it applies)
and its limitations (scenarios that may result in
limited parallelism or suboptimal solutions).

\subsection{Generality}\label{sec:general}

\sysname is applicable to a wide range of real-world
resource allocation problems.
Table~\ref{tab:comparison} summarizes these problems that appear in
recent literature such as OSDI, SOSP, NSDI, and SIGCOMM.
Below, we explain why \sysname is applicable.

\begin{itemize}[noitemsep,topsep=0pt,leftmargin=*]

\item \parai{Variables.} As an ADMM-based framework, \sysname natively supports continuous (real-valued) allocation variables. Empirical evidence\cite{admmint} and theoretical proofs~\cite{admmnonconvex, admmnonconvex2, admmnonconvex3} on ADMM further indicate that \sysname should effectively handle boolean and integer variables by projecting real-valued solutions onto the appropriate domains during iterations.

\item \parai{Objective.} Most objectives in real systems are defined as the sum, max,
or min of utility functions over the individual allocation of
resources and demands. They also exhibit convexity. Examples include weighted sums~\cite{narayanan2020heterogeneity, xu2023teal}, logarithmic utilities~\cite{agrawal2022allocation}, and quadratic costs~\cite{yang2023skypilot, werner2021pricing}.

\item \parai{Constraints.} \sysname assumes linear constraints on resources and
demands---a condition that holds across all the surveyed use cases.
These constraints are commonly expressed at the per-resource or per-demand level.
Even for constraints that do span multiple resources or demands, \sysname can turn them
into disjoint constraint groups and apply decomposition at the group level
(at the cost of reduced parallelism).
\end{itemize} 

Unlike POP~\cite{narayanan2021solving}, \sysname works with any workload, not only ``granular'' ones.
In POP, the demands in each subproblem can only access a subset of resources,
assuming each demand only requests a small fraction of interchangeable resources such that a resource subset suffices.
This granular assumption can be unrealistic in practice, due to uneven demand distributions~\cite{jeon2019analysis, wang2021examination} or specific resources that each demand is allowed to utilize~\cite{FGD2023, kumar2018semi}.
In contrast, each subproblem in \sysname can still access all the resources or demands, resulting in generalization to various workload distributions.

\subsection{Limitations}\label{sec:exception}

While \sysname applies to a variety of resource allocation tasks, 
it might not always be the optimal approach for all scenarios.

\parab{Limited parallelism.} \sysname's parallelism relies on its ability to decompose
the problem into independent subproblems. When such decomposition
is not fully achievable, the degree of parallelism diminishes.
Although this limitation does not manifest in the set of real-world
problems we surveyed, 
we come up with some examples below.

\begin{itemize}[noitemsep,topsep=0pt,leftmargin=*]
\item
\parai{Non-separable objectives.}
An objective is non-separable when utility depends on the interaction
between different resources or demands, e.g., if a job attains higher throughput
when it has exclusive access to a GPU compared with when sharing it with other jobs.
\item
\parai{Non-separable constraints.} A constraint is non-separable when it
spans multiple resources and multiple demands simultaneously. For example,
this occurs when the jobs submitted by each user are collectively limited by
a user-specific GPU hour quota.
To restore decomposability, \sysname has to treat all jobs (the original demands)
from the same user as a single aggregated demand.
However, this aggregation reduces the granularity of parallelism,
from the job level to the user level.

\end{itemize}

\parab{Suboptimal solutions.} \sysname may fail to reach the optimal solution
when the allocation problem deviates from the assumptions underlying standard ADMM.

\begin{itemize}[noitemsep,topsep=0pt,leftmargin=*]

\item 
\parai{Non-convex problems.}
Although ADMM has demonstrated empirical success on various
non-convex problems~\cite{admmint, admmempirical, admmempirical2},
its theoretical convergence is guaranteed only under specific conditions~\cite{admmnonconvex, admmnonconvex2, admmnonconvex3}.
Consequently, \sysname may converge to suboptimal solutions in non-convex settings.

\item 
\parai{Nonlinear constraints.} When a constraint is nonlinear, \sysname has to reformulate
it with an indicator function in the augmented Lagrangian. The indicator function returns zero if the constraint is satisfied and infinity otherwise. However, this approach destroys the convexity of the objective, resulting in the aforementioned non-convex problems.

\item 
\parai{Higher allocation dimension.}
\sysname assumes a two-dimensional allocation matrix that maps resources
to demands, leading to a two-block ADMM formulation.
Introducing additional dimensions such as a temporal dimension~\cite{timealloc} will increase the number of variable blocks in ADMM and thus might lead to convergence issues~\cite{admm3blocks}.
\end{itemize}

\section{Case Studies}
\label{sec:cases}

In this section, we present case studies of \sysname applied to three representative resource allocation problems~\cite{narayanan2021solving}: cluster scheduling, traffic engineering, and load balancing.

\vspace{-5pt}
\subsection{Cluster Scheduling}
\label{sec:cluster-scheduling}

Cluster scheduling problems~\cite{xiao2018gandiva, narayanan2020heterogeneity, narayanan2021solving, namyar2023solving} involve allocating computing jobs to different resource types in heterogeneous clusters.
The job scheduling example introduced earlier is a simplified version of this problem.

\parab{Max-min allocation.}
This variant aims to distribute the heterogeneous cluster resources fairly
among jobs such that the minimum throughput is maximized.

\begin{itemize}[itemsep=0pt,topsep=0pt,leftmargin=*]
\item \parai{Variables.} Jobs can be time-sliced across available resource types~\cite{narayanan2020heterogeneity, narayanan2021solving}. Thus, the allocation plan
is a matrix $x \in [0, 1]^{n \times m}$, where $x_{ij}$ is the fraction of time
(optimization interval) that each job $j$ spends on each resource type $i$.
\item \parai{Objective.} Given an allocation matrix $x$, we adopt a normalized effective throughput of job $j$ from POP~\cite{namyar2022minding}, which is a linear function
of $x_{*j}$. The objective is to maximize the minimum normalized effective throughput:
\[ \maximize_{x} \min_j~\text{throughput}(j, x_{*j}). \]
\vspace{-15pt}
\item \parai{Resource constraints.} For each computing resource type $i$, the total amount of resources requested by all jobs cannot exceed the corresponding capacity:
\[ \sum_j x_{ij}z_j \leq \text{capacity}_i, \quad \forall i,\]
where $z_j$ is the amount of resources requested by job $j$.
\item \parai{Demand constraints.} For each job $j$, the sum of all fractions of time must not exceed 1 naturally:
\[ \sum_i x_{ij} \leq 1, \quad \forall j.\]
\vspace{-8pt}
\end{itemize}

\parab{Proportional fairness.} The second variant of cluster scheduling maximizes the overall resource utilization while ensuring minimum service for each job. This variant shares the same variables and constraints as above,
except for the objective:
\begin{itemize}[noitemsep,topsep=0pt,leftmargin=*]
\item \parai{Objective.} Proportional fairness aims to maximize the sum of log utilities of each job:
\[ \maximize_{x} \sum_j \log(\text{throughput}(j, x_{*j})). \]
\vspace{-10pt}
\end{itemize}

\parab{Applying \sysname.} 
Both variants of cluster scheduling exhibit the separable structure required by \sysname. 
Their objectives to maximize are a sum or min of utilities for allocating each job,
and a linear constraint is applied to each job and each computing resource.
Thus, \sysname can decompose the problem into $n$ per-job allocations and $m$ per-demand allocations.

\subsection{Traffic Engineering}

Traffic engineering problems~\cite{hong2013achieving, valadarsky2017learning, narayanan2021solving, namyar2023solving, xu2023teal} allocate traffic demands between datacenters onto the network links in a wide-area network (WAN) topology.

\parab{Maximizing total flow.} This problem variant focuses on maximizing
the total flow allocation.

\begin{itemize}[itemsep=0pt,topsep=0pt,leftmargin=*]
\item \parai{Variables.} Let $x_{(u, v)(s,t)} \geq 0$ denote the amount of flow
from a datacenter pair $(s, t)$ assigned to a network link $(u, v)$.
In common path-based traffic engineering, flows between each node pair $(s, t)$ are allocated only over links along pre-configured paths $P_{(s,t)}$.
\item \parai{Objective.} For each node pair $(s, t)$, the inflow it incurs to node $v$ is computed as: $\text{inflow}(v, (s,t)) = \sum_{(u, v) \in P_{s,t}} x_{(u, v)(s,t)}$,
and outflow as: $\text{outflow}(v, (s,t)) = \sum_{(v, u) \in P_{s,t}} x_{(v, u)(s,t)}$.
Consequently, the total flow allocated for the node pair $(s, t)$ is captured by $\text{inflow}(t, (s,t))$, the total inflow toward the destination $t$. Therefore, the objective of maximizing the total flow between all node pairs can be expressed as:
\[\maximize_x \sum_{(s,t)} \text{inflow}(t, (s,t)).\]
\vspace{-12pt}
\item \parai{Resource constraints.} For each link $(u, v)$, the cumulative flow from all node pairs traversing it must not exceed its link capacity $c_{u,v}$:
\[ \sum_{(s,t)} x_{(u, v)(s,t)} \leq c_{u, v}, \quad \forall (u, v) \in E.\]
\vspace{-12pt}
\item \parai{Demand constraints.} For each node pair $(s, t)$, the total flow allocated cannot exceed the total demand. Additionally, for all nodes other than $s$ and $t$, the inflow and outflow must be equal, ensuring no flow ``black hole'' along the paths:
\[\begin{array}{l}
0 \leq \text{inflow}(t, (s,t)) \leq d_{s,t}, \\
\text{inflow}(v, (s,t)) = \text{outflow}(v, (s,t)),  \forall v \neq s, t, \\
\end{array}
\quad \forall (s, t).
\]
\end{itemize}

\parab{Minimizing max link utilization.} This variant aims to evenly distribute traffic demands across network links to avoid overloading any single link. 
The utilization of link is commonly defined as the (uncapped) ratio of total
flow assigned to a link to the link's capacity,
although it cannot actually exceed 100\% in practice.
This formulation shares the same variables and constraints as above, with a revised
objective:
\begin{itemize}[noitemsep,topsep=0pt,leftmargin=*]
\item \parai{Objective.} The goal is to minimize the max link utilization across the network:
\[
\minimize_x \max_{(u, v) \in E} \frac{\sum_{(s,t)} x_{(u, v)(s,t)}}{c_{u, v}}.
\]
\vspace{-1em}
\end{itemize}

\parab{Applying \sysname.} The traffic engineering problem aligns with the separable structure of \sysname: the objectives are expressed as either a sum over
per-demand utilities or as a maximum over per-link utilities, and
linear constraints are applied to each demand and each link.
Therefore, \sysname can decompose the problem into $|E|$ per-link allocations and $|V|^2$ per-demand allocations for parallelism. In practice, due to the large overhead of maintaining $|V|^2$ per-demand subproblems, we group these subproblems by their sources, reducing the total number of subproblems to just $|V|$.

\subsection{Load Balancing} \label{sec:load_balancing_form}

Load balancing problems~\cite{curino2011workload, serafini2014accordion, taft2014store, narayanan2021solving} allocate data shards among storage servers to scale out query loads in a distributed store. Load balancing is a non-convex problem with integer variables, wherein \sysname can provide a high-quality solution empirically, as discussed in \S\ref{sec:exception}.

\parab{Minimizing shard movements.} This problem aims to minimize shard movements across servers during load changes while keeping the load nearly balanced on each server.
\begin{itemize}[itemsep=0pt,topsep=0pt,leftmargin=*]
\item \parai{Variables.}
The allocation plan is encoded in a matrix $x \in [0, 1]^{n \times m}$, where $x_{ij}$ is the fraction of data shard $j$ assigned to server $i$. 
Additionally, a binary matrix $x'$ captures shard placement, where $x'_{ij} = 1$ if the data shard $j$ is located on server $i$ (i.e., when $x_{ij} > 0$);
otherwise, $x'_{ij} = 0$.
\item \parai{Objective.} The movement of data shard $j$ allocated to server $i$ can be represent as $(1 - T_{ij}) x'_{ij}f_i$, where $T_{ij}$ denotes the initial shard placement,
and $f_j$ denotes the memory footprint of data shard $j$. The objective of minimizing total shard movement can be expressed as:
\[ \minimize_{x} \sum_j \sum_i (1 - T_{ij}) x'_{ij} f_i. \]
\vspace{-15pt}
\item \parai{Resource constraints.} For each server $i$, the total query load must be close to the average query load $L$. Meanwhile,
the total memory usage must be within the memory capacity:
\[    
\begin{array}{l}
L - \epsilon \leq \sum_j x_{ij} l_j \leq L + \epsilon, \\
\sum_j x^\prime_{ij} f_j \leq \text{memory}_i,
\end{array}
\quad \forall i,
\]
where $l_j$ is the query load on data shard $j$.
\item \parai{Demand constraints.} Each data shard $j$ must be entirely allocated
across servers:
\[ \sum_{i}x_{ij} = 1, \quad \forall j. \]
\end{itemize}

\parab{Applying \sysname.} 
The load balancing problem also demonstrates a
separable structure: the objective is the sum of allocation quality per data
shard, while linear constraints are applied independently to each server and each shard. This allows \sysname to decompose the problem into $n$ per-server allocations
and $m$ per-shard allocations.

\section{Implementation of \sysname}
\label{sec:implementation}

We have implemented \sysname as a Python package
and uploaded it to PyPI for easy installation:
\texttt{pip install dede}.

Listing~\ref{algo:dede} showcases a basic resource allocation example
using \sysname. 
Line~5 creates an $N\times M$ non-negative allocation matrix \texttt{x},
where \texttt{x[i,j]} represents the fraction of time that demand \texttt{j}
runs on resource \texttt{i}.
Lines~12--13 define a constraint for each resource using parameters,
and Lines~14--15 define a constraint for each demand.
The objective, defined in Line 18, is to maximize the sum of entries in \texttt{x}.
Finally, the problem is constructed in Lines~21--22 and solved in Line~23.

Built on top of the popular optimization modeling language
\texttt{cvxpy}~\cite{cvxpy1, cvxpy2}, \sysname inherits most of
its syntax and APIs, such as \texttt{Variable()}, \texttt{Parameter()},
and \texttt{Maximize()}.
A notable distinction is that \sysname requires users to explicitly
separate resource constraints and demands constraints
when initializing a problem (Line~22).
As a parallel optimization tool, \sysname allows users to configure the number of CPUs~(Line~23); by default, all available cores are used.

Internally, \sysname solves the optimization in three stages:
\begin{itemize}[noitemsep,topsep=0pt,leftmargin=*]
\item \parai{Problem parsing.} \sysname transforms all inequality constraints into equivalent equality constraints through non-negative slack variables. E.g., \texttt{x[:,j].sum() <= 1} is converted to \texttt{x[:,j].sum() + sj == 1}
with a slack variable \texttt{sj}.
Then slack variables are treated just like other variables.
\item \parai{Problem building.} \sysname organizes resource constraints into
disjoint per-resource groups and demand constraints into disjoint per-demand groups.
For each group, a per-resource or per-demand subproblem 
is constructed with \texttt{cvxpy} following Equations~\ref{eq:resource_subproblem}
and~\ref{eq:demand_subproblem}.
\item \parai{Problem solving.} \sysname solves these subproblems in parallel using multiple CPU cores. During ADMM iterations, only the parameters are updated, avoiding the overhead of rebuilding problems in
\texttt{cvxpy}. Similarly, for the same problem with varying resources and demands,
only the relevant parameters are updated.
\end{itemize}

\begin{lstlisting}[style=myStyle, captionpos=b, float=t!, 
    caption={Resource allocation example with \sysname package.},
    label={algo:dede}, 
    abovecaptionskip=0.1in, belowcaptionskip=-0.2in]
import numpy as np
import dede as dd

# Create allocation variables
x = dd.Variable((N, M), nonneg=True)

# Create parameters
param = dd.Parameter(
    N, value=np.random.uniform(0, 1, N))

# Create constraints
resource_constrs = [
    x[i,:].sum() <= param[i] for i in range(N)]
demand_constrs = [
    x[:,j].sum() <= 1 for j in range(M)]

# Create an objective
obj = dd.Maximize(x.sum())

# Construct and solve the problem
prob = dd.Problem(
    obj, resource_constrs, demand_constrs)
prob.solve(num_cpus=64, solver=dd.ECOS)
\end{lstlisting}

We use Ray~\cite{moritz2018ray} to parallelize our computation across multiple cores,
as Python's native multithreading model is constrained by the (notorious)
global interpreter lock (GIL). 
Ray circumvents this issue by managing multiple Python
interpreter processes, effectively bypassing the GIL.
Moreover, Ray offers robust support for inter-process communication, sparing us
from manually implementing such mechanisms using Python's lower-level
multiprocessing primitives.

\section{Evaluation}
\label{sec:eval}

We answer the following questions in the evaluation:
\begin{enumerate}[noitemsep,topsep=0pt,leftmargin=*]
\item How does \sysname compare with state-of-the-art approaches in allocation quality and computation time?~(\S\ref{sec:eval_basic})
\item How do \sysname and baseline approaches react to changes in problem granularity, temporal fluctuation, spatial redistribution, and link failures?~(\S\ref{sec:eval_change})
\item What is the individual contribution of each design component in \sysname?~(\S\ref{sec:eval_micro})
\end{enumerate}

\smallskip
Our evaluation compares four approaches:
\begin{itemize}[noitemsep,topsep=0pt,leftmargin=*]
\item
\parai{Exact sol.} This baseline solves the original resource allocation problem using commercial solvers. We adopt the implementation from POP, which utilizes \texttt{cvxpy}~\cite{cvxpy1, cvxpy2}, Gurobi~\cite{gurobi}, and CPLEX~\cite{cplex} for cluster engineering, traffic engineering, and load balancing, respectively.
\item
\parai{POP.} POP-$k$~\cite{narayanan2021solving} randomly splits the problem into $k$ smaller subproblems, applies commercial solvers to each, and coalesces the resulting $k$ sub-allocations into a global allocation. However, POP only \textit{simulates} the parallel execution by solving subproblems sequentially
and calculating the parallel solving time mathematically.
\item
\parai{\sysname.} \sysname is our truly parallel implementation (\S\ref{sec:implementation}). By default, the solution
from the previous optimization interval
is used to warm-start the subsequent interval.
\item
\parai{\sysnamestar.} For a fair comparison with POP,
we create a variant of \sysname, denoted as \sysnamestar, following POP's simulation
methodology. \sysnamestar solves subproblems sequentially and estimates
the parallel solving time mathematically. 
\end{itemize}

We also consider domain-specific approaches apart from these four approaches. 
In cluster scheduling and load balancing, we evaluate the greedy heuristic algorithms, Gandiva~\cite{xiao2018gandiva} and E-Store~\cite{taft2014store}, respectively.
In traffic engineering, we evaluate a demand-pinning approach~(``Pinning'') similar to a prior work~\cite{namyar2022minding}, where the top 10\% of demands are allocated using optimization engines and the rest are assigned to shortest paths. We also evaluate Teal~\cite{xu2023teal}, a learning-accelerated traffic engineering approach that
achieves massive parallelism using a GPU.

All approaches are evaluated in terms of allocation quality and computation time.
The computation time is measured using 64 CPU cores (2$\times$ Intel Xeon Gold 6142),
with an additional GPU~(Nvidia Titan RTX) available for Teal. 

\subsection{\sysname vs. the State of the Art}\label{sec:eval_basic}

In this section, we compare \sysname against state-of-the-art approaches on resource allocation problems from \S\ref{sec:cases}.

\subsubsection{Cluster Scheduling}
\label{sec:eval_cluster_scheduling}

We stress-test \sysname's scalability in a large, heterogeneous cluster scheduling
environment, where jobs are allocated in a time-sliced manner across
456 different types of compute resources, e.g., GPU/CPU instances that
vary in vendor, generation, memory capacity, and other specifications.
The quantity of each resource type is randomly drawn from
$\{8, 16, 24, \ldots, 64\}$, resulting in a total of 16,520 instances.
We utilize the simulator in Gavel~\cite{narayanan2020heterogeneity}
and model job arrivals as a Poisson process with an average inter-arrival
of 100 seconds.
We update job allocation decisions every 6 minutes, repeating this
process for 200 scheduling rounds (20 hours).
For each job, the number of requested resource instances per type is drawn from
$\{1, 2, 4, 8, 16, 32\}$,
and following prior work~\cite{FGD2023},
33\% of the jobs are restricted to specific resource types.
Job throughput is derived from relevant benchmarks~\cite{ignatov2019ai, lambda, EpochNotableHardwares2024}.
Please refer to Appendix~\ref{sec:setup_cluster_scheduling}
for more details on this setup.

\begin{figure}[t]
    \centering
    \includegraphics[height=110pt]{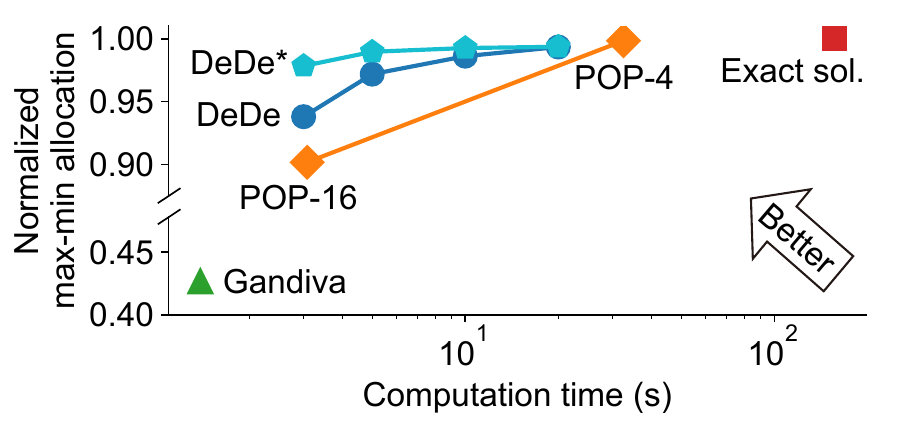}
    \vspace{-5pt}
    \caption{Results of the cluster scheduling variant that maximizes the minimum job throughput
    (max-min allocation).}
    \vspace{-5pt}
    \label{fig:basic_cs}
\end{figure}

\parab{Max-min allocation.} Figure~\ref{fig:basic_cs} compares \sysname against other approaches in terms of max-min allocation. \sysname quickly attains a near-optimal max-min allocation of 0.94 in just 3 seconds and 0.99 within 10 seconds, demonstrating a better balance between max-min allocation and computation time compared with Exact sol., Gandiva, and POP. 
In contrast, Exact sol. experiences a significantly slower computation time of 156 seconds due to its sequential nature. The greedy heuristic, Gandiva, performs poorly, achieving only a max-min allocation of 0.43 (in 1.4 seconds).
POP-4 provides a comparable tradeoff between time and accuracy to \sysname, but is 1.6$\times$ slower. 
The faster variant, POP-16, splits the problem into 16 smaller subproblems and achieves a lower max-min allocation of 0.9 (in 3.1 seconds),
since smaller subproblems limit the ability to select optimal GPU/CPU assignments.

\sysnamestar, representing the same algorithm under idealized conditions
similar to POP's simulation, achieves a normalized max-min allocation of 0.99 and is 3.3$\times$ faster than \sysname in Figure~\ref{fig:basic_cs}. 
The reason is that \sysnamestar executes each iteration more quickly and completes more iterations within a given time. The discrepancy in iteration time between \sysname and \sysnamestar arises from three factors: 
(1) \sysname measures the end-to-end time, including problem solving, compilation, solution unpacking, and other associated tasks,
while \sysnamestar only accounts for the core solving time;
(2) \sysname's parallel implementation incurs overhead from cache contention, which slows computation; 
(3) In \sysname, each subproblem is statically pre-assigned to one of the processes, making it susceptible to straggler delays. In contrast, \sysnamestar assumes perfect dynamic scheduling.

\begin{figure}[t]
    \centering
    \includegraphics[height=110pt]{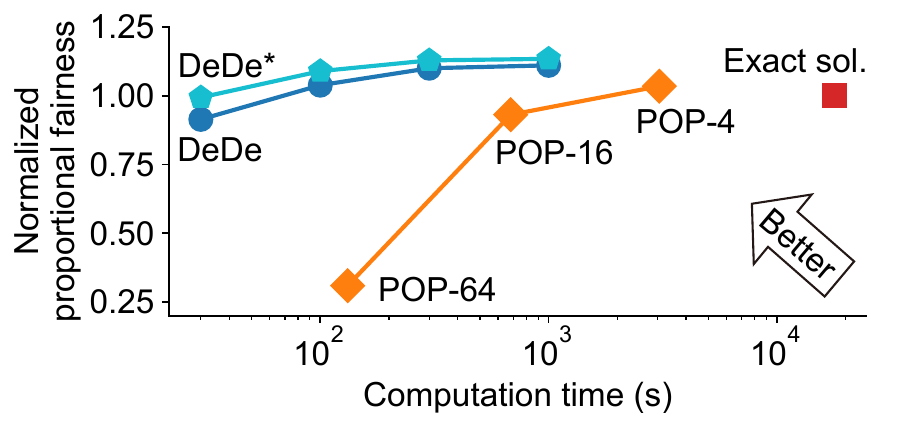}
    \vspace{-5pt}
    \caption{Results of the cluster scheduling variant that maximizes proportional
    fairness among job throughputs.}
    \label{fig:basic_log_cs}
    \vspace{-5pt}
\end{figure}

\parab{Proportional fairness.} Figure~\ref{fig:basic_log_cs} compares \sysname against other approaches in maximizing proportional fairness. 
This problem is particularly challenging due to the logarithmic form of
the objective function.
Exact sol., which uses the SCS solver in \texttt{cvxpy}, fails to reach
optimality even after 5 hours of computation. As a result, several methods report
normalized fairness scores that exceed 1 when normalized relative to Exact sol.
Both \sysname and \sysnamestar achieve greater speedups than POP by effectively breaking the problem into smaller and easier subproblems. Notably, \sysname and \sysnamestar both reach a normalized proportional fairness of over 1 within 100 seconds. 
In contrast, POP-4 and POP-16 require 3,053 seconds and 682 seconds, respectively, to achieve comparable allocation quality, while POP-64 yields a significantly lower fairness score of only 0.31. 

\subsubsection{Traffic Engineering}
\label{sec:eval_traffic_engineering}

\begin{figure}[t]
    \centering
    \includegraphics[height=115pt]{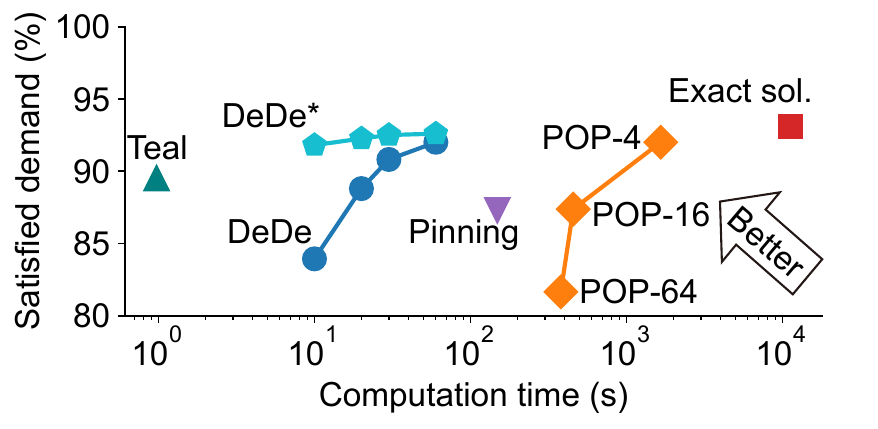}
    \vspace{-5pt}
    \caption{Results of the traffic engineering problem variant that maximizes the total flow.}
    \label{fig:basic_te}
    \vspace{-5pt}
\end{figure}

We adopt the evaluation setup in Teal~\cite{xu2023teal},
with a focus on its largest topology---a 1,739-node network
adapted from an internet topology for WAN purposes.
The traffic matrices are collected from the production WAN of a global
cloud provider and mapped onto the above topology.

\parab{Maximize total flow.}
Figure~\ref{fig:basic_te} compares \sysname to the baselines in maximizing total flow. 
\sysname achieves 90.8\% satisfied demand within 30 seconds and 92\% within 60 seconds, showing a superior trade-off between satisfied demand and computation time compared with Exact sol., POP, and Pinning. Its idealized variant, \sysnamestar, achieves a satisfied demand that is 8\% higher than \sysname within 10 seconds. While POP-4 also reaches 92\% satisfied demand, its average computation time is significantly longer, at 1,658 seconds. POP-16 and POP-64 improve parallelism by dividing the workload into more subproblems, reducing computation times to 456 seconds and 380 seconds, respectively. Note that the speedup of POP-64 is limited compared with POP-16 as the number of cores per subproblem drops from 4 to 1. This increased parallelism also comes at a cost, with satisfied demand decreasing to 87.4\% for POP-16 and 81.6\% for POP-64.
Pinning focuses on top demands to reduce problem size but remains constrained by the sequential nature of its optimization solvers. As a result, it achieves a modest speedup, requiring 149 seconds to reach 87.3\% satisfied demand.
Teal, a learning-accelerated approach, leverages massive GPU parallelism to achieve 89\% satisfied demand in only 1 second. However, it relies on a highly customized machine-learning framework, demanding significant human effort for design and training.

\begin{figure}[t]
    \centering
    \includegraphics[height=115pt]{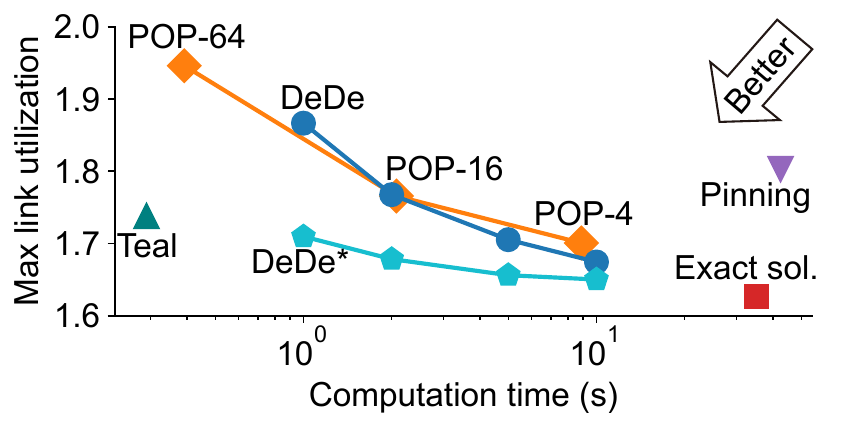}
    \vspace{-5pt}
    \caption{Results of the traffic engineering problem variant
that minimizes the max link utilization.}
    \label{fig:basic_mlu_te}
    \vspace{-5pt}
\end{figure}
\begin{figure}[t]
    \centering
    \includegraphics[height=115pt]{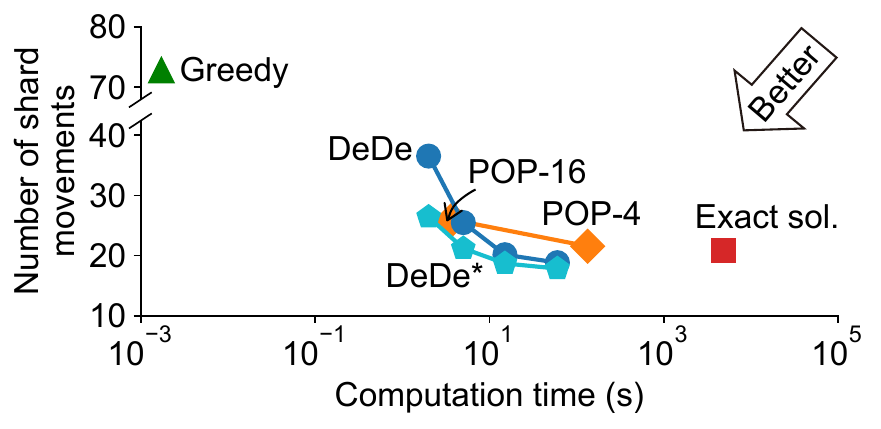}
    \vspace{-5pt}
    \caption{Results of the load balancing problem where the objective is to
    minimize shard movements.}
    \label{fig:basic_lb}
    \vspace{-5pt}
\end{figure}

\parab{Min-max link utilization.} Figure~\ref{fig:basic_mlu_te} compares \sysname with other approaches in minimizing maximum link utilization. Here, the link utilization metric serves as proxy for network congestion and is allowed to exceed 100\% during optimization. \sysname attains a maximum link utilization of 1.67 within 10 seconds and 1.71 within 5 seconds.
Exact sol. reaches 1.63 with a longer computation time of 35 seconds. This is faster than maximizing total flow because minimizing max link utilization is computationally less
challenging and thus requires fewer iterations.
Pinning takes 42.5 seconds, making it slower than \sysname and slightly slower than Exact sol., as it requires additional time to rebuild the problem when the top 10\% of node pairs change. 
POP's variants, POP-4, POP-16, and POP-64, achieve a maximum link utilization of 1.70, 1.77, and 1.95 within 9, 2, and 0.4 seconds on average, respectively. 
Although POP-16 is faster than \sysname, a fairer comparison between POP-16 and \sysnamestar, both simulating parallelism, reveals that \sysnamestar achieves a comparable computation time to POP-16.
The domain-customized Teal takes only 0.3 seconds on a GPU to attain a
maximum link utilization of 1.74.

\subsubsection{Load Balancing}

\begin{figure*}[t]
\hspace{10pt}
\begin{subfigure}[t]{0.36\textwidth}
\includegraphics[width=0.85\columnwidth]{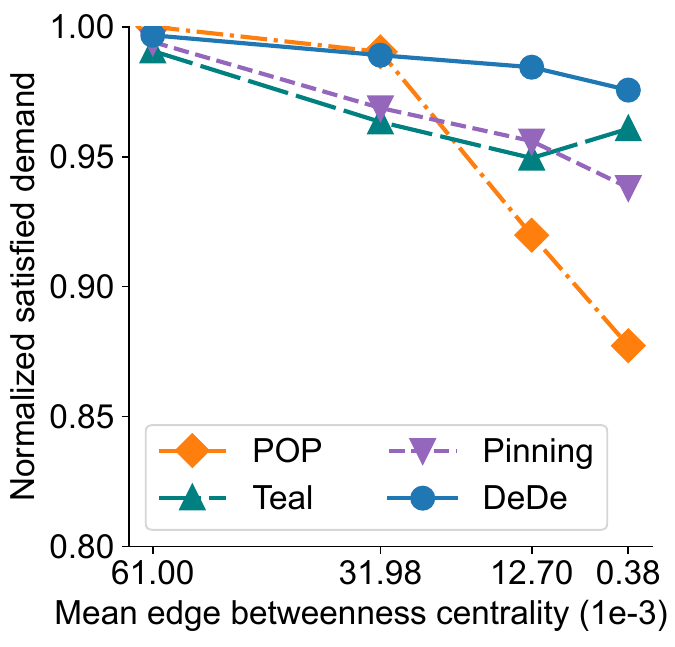}
\caption{Granularity changes.}
\label{fig:centrality}
\end{subfigure}
\hspace{-30pt}
\begin{subfigure}[t]{0.36\textwidth}
\includegraphics[width=0.85\columnwidth]{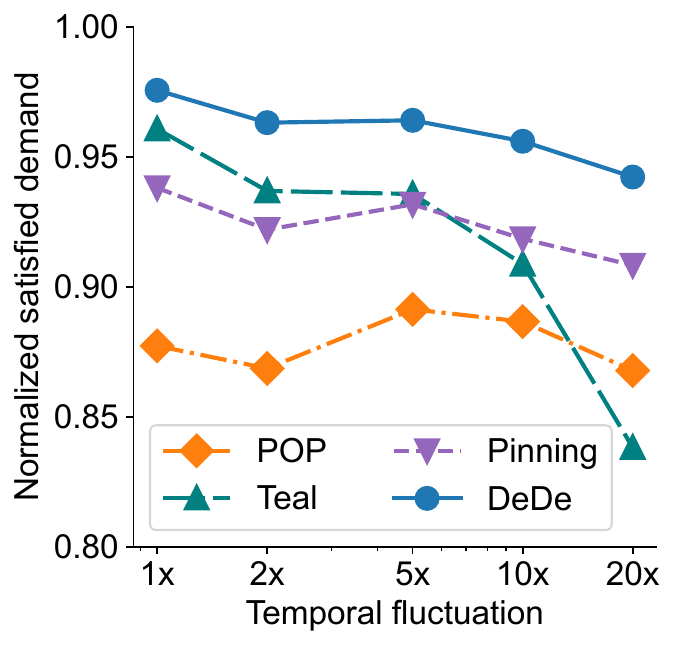}
\caption{Temporal fluctuation.}
\label{fig:temporal}
\end{subfigure}
\hspace{-30pt}
\begin{subfigure}[t]{0.36\textwidth}
\includegraphics[width=0.85\columnwidth]{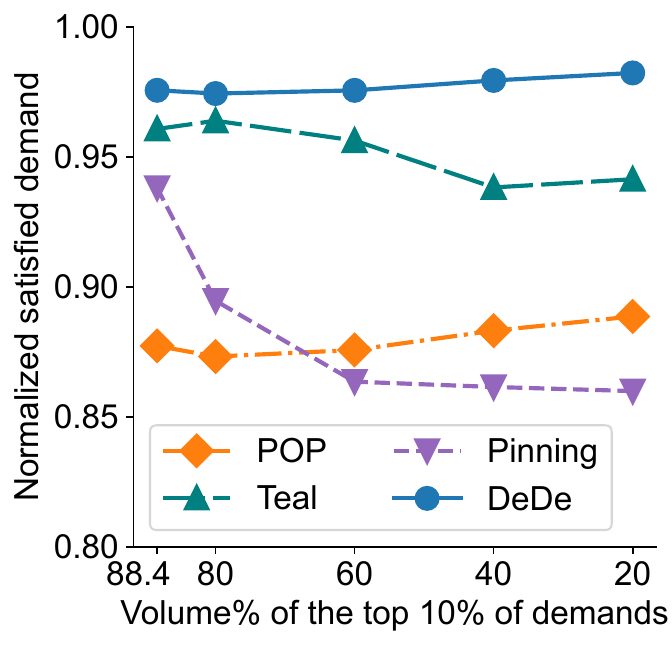}
\caption{Spatial redistribution.}
\label{fig:spatial}
\end{subfigure}
\vspace{-3pt}
\caption{Normalized satisfied demand in traffic engineering against granularity, temporal, and spatial changes.}
\label{fig:changes}
\vspace{-5pt}
\end{figure*}

We follow the load balancing settings in POP~\cite{narayanan2021solving},
except for increasing the scale to 2,048 data shards distributed across 256 servers.
In each round, a new shard-to-server allocation is computed based on 
updated query loads. 
We simulate for 100 rounds, using the first 20 as a warm-up.
To ensure the problem's feasibility, we set the tolerance parameter
$\epsilon$ to 0.1.

\parab{Minimize shard movements.}
Figure~\ref{fig:basic_lb} compares \sysname with baselines in minimizing shard movements.
\sysname achieves an average of 20.1 shard movements in 15 seconds, better than POP-4's 21.5 shard movements within 133 seconds. In faster scenarios, \sysname achieves 25.4 shard movements in 5 seconds, while POP-16 achieves an average movement of 26 in a slighter less time of 3.7 seconds.
For a fair comparison, \sysnamestar consistently achieves around 16\% fewer shard movements than POP, given a similar time limit.
Exact sol. requires 20.9 shard movements on average, but is significantly slower with an average of 4,820 seconds. This slower computation is due to the NP-hard nature of solving the mixed-integer linear program underlying load balancing. Notably, Exact sol. does not produce the minimum possible shard movements because optimizing each scheduling round independently does not guarantee an optimal solution across the entire series.
At the opposite end of the time-accuracy trade-off, the greedy heuristic provides a rapid but less accurate solution, completing in just 2 milliseconds but averaging 73 shard movements after naively fixing its constraint violations.

\begin{figure*}[t]
\hspace{5pt}
\begin{subfigure}[t]{0.36\textwidth}
\includegraphics[width=0.85\columnwidth]{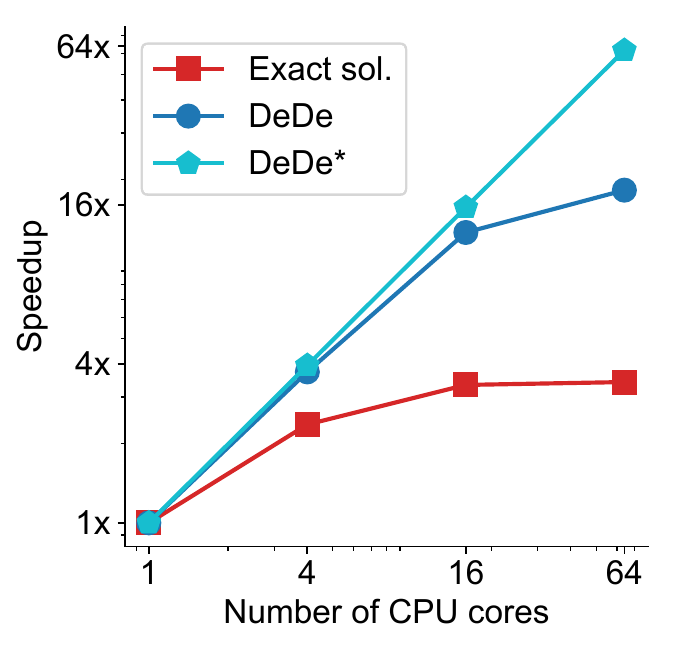}
\caption{Speedup when varying CPU cores.}
\label{fig:cpu_time}
\end{subfigure}
\hspace{-30pt}
\begin{subfigure}[t]{0.36\textwidth}
\includegraphics[width=0.85\columnwidth]{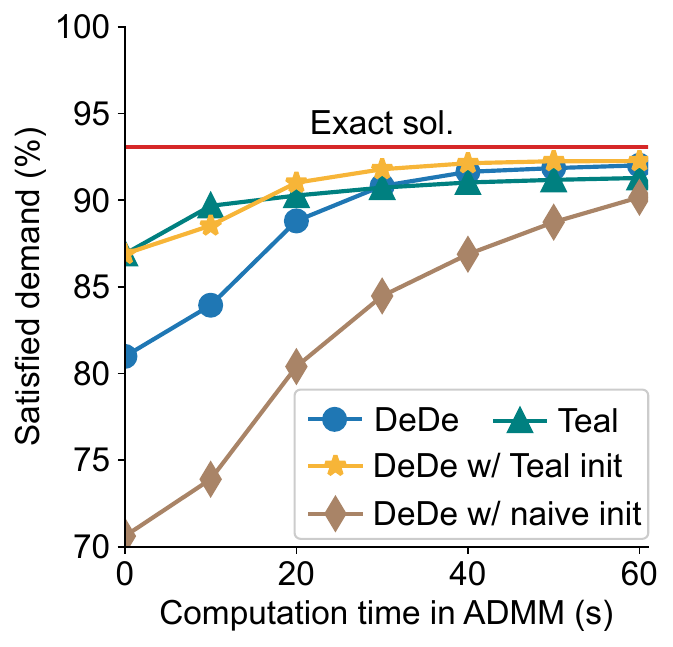}
\caption{Convergence rate.}
\label{fig:iter}
\end{subfigure}
\hspace{-30pt}
\begin{subfigure}[t]{0.36\textwidth}
\includegraphics[width=0.85\columnwidth]{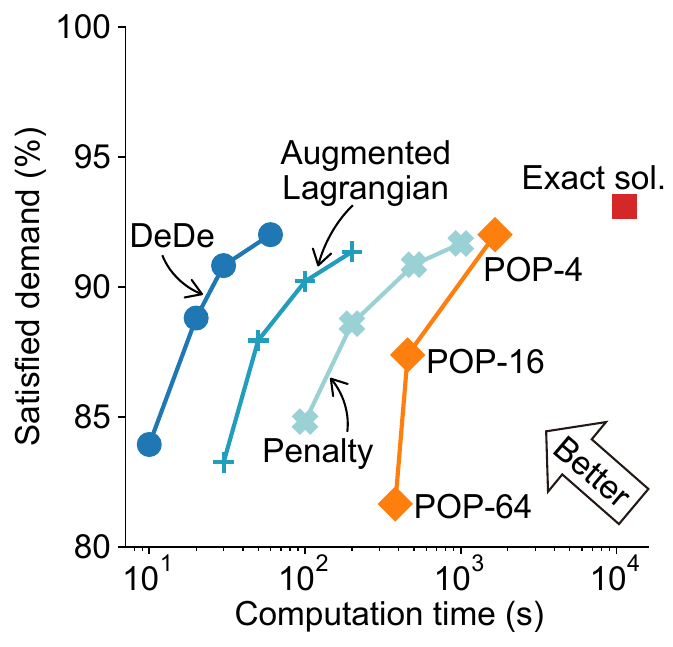}
\caption{Comparing different optimization methods.}
\label{fig:decouple}
\end{subfigure}
\vspace{-2pt}
\caption{Micro-benchmarks with the objective of maximizing total flow in traffic engineering.}
\label{fig:microbm}
\vspace{-5pt}
\end{figure*}

\begin{figure}[t]
    \centering
    \includegraphics[width=0.99\columnwidth]{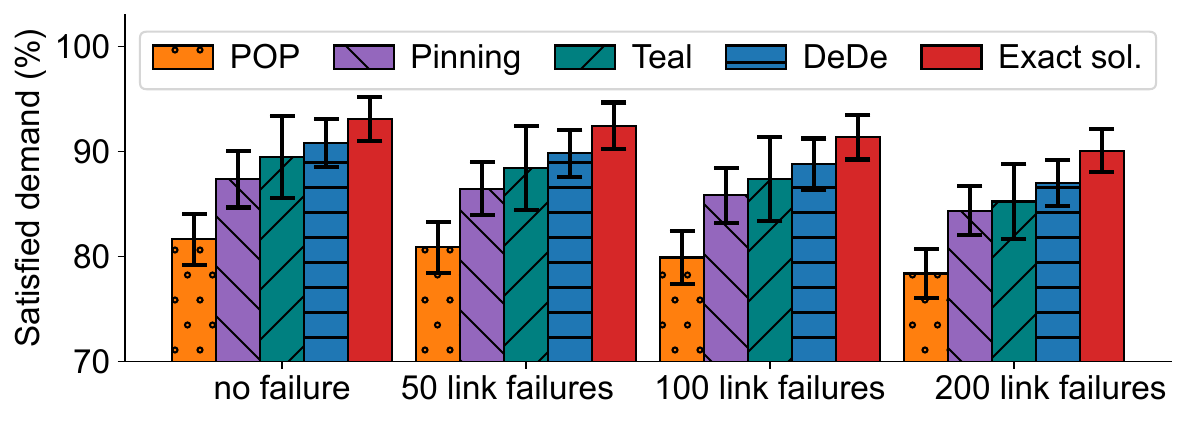}
    \vspace{-5pt}
    \caption{Satisfied demand of traffic engineering under zero, 50, 100, or 200 link failures after recomputing flow allocation.}
    \label{fig:failures}
    \vspace{-10pt}
\end{figure}

\subsection{Robustness}\label{sec:eval_change}
Using traffic engineering as an example, we evaluate the robustness of \sysname and other baselines against changes in problem granularity, temporal dynamics,
spatial redistribution, and link failures. In these experiments, \sysname is run for 30 seconds, and we normalize the satisfied traffic demand by the optimal one achieved by
Exact sol.

\parab{Changes of problem granularity.} The granular problem structure required by POP, where each demand only requests a small fraction of \textit{interchangeable} resources,
can be impractical.
In traffic engineering, for example, demands are constrained by the link capacity on specific pre-configured paths. Similarly, in cluster scheduling, 33\% of GPU tasks in production clusters are limited to specific GPU types~\cite{FGD2023}. 
In Figure~\ref{fig:centrality}, we present results on varying levels
of problem granularity using traffic engineering as an example.
To quantify resource interchangeability, we use the mean edge betweenness centrality~\cite{brandes2001faster, brandes2008variants}, which measures the average percentage of demands served by a given edge.

As the mean edge betweenness centrality decreases, all approaches experience a decline in normalized satisfied demand, ranging from 0.8--2.1\% for \sysname, 2.8--4.2\% for Teal, 2.5--5.6\% for Pinning, and 1--12.3\% for POP. 
However, the maximum decrease in POP is 5.9$\times$ greater than \sysname because POP's subproblems struggle to find alternative resources within their restricted subsets, whereas \sysname's subproblems retain access to the full set of resources.
Teal exhibits a slight 1.3\% increase in satisfied demand when the mean edge betweenness centrality is minimal. This is because the reduced edge selection simplifies allocation decisions.

\parab{Temporal dynamics.} Next, we introduce temporal fluctuations~\cite{xu2023teal} to the traffic matrices for robustness assessment. For each demand, we calculate the variance $\sigma^2$ in its changes between consecutive time slots and create a new normal distribution $N(0, k\sigma^2)$. We choose the values of $k$ to be 2, 5, 10, and 20. Next, we randomly draw a sample from the newly created normal distribution and add it to each demand in every time slot, simulating temporal fluctuations.

Figure~\ref{fig:temporal} demonstrates that \sysname effectively manages temporal fluctuations, with changes in normalized satisfied demand limited to 1.3--3.5\%. In contrast, Teal suffers a significant decrease of 12.8\%, as it lacks prior exposure to the distribution of these fluctuating demands during training. Pinning and POP experience variations in normalized satisfied demand ranging from 0.6--3\% and 1--1.6\%, respectively. This is because randomness may occasionally favor splitting a problem into subproblems or focusing on top demands exclusively, while at other times, it does not.

\parab{Spatial redistribution.} To evaluate robustness against spatial redistribution, we reassign traffic demands among node pairs. Specifically, we adjust the top 10\% of demands, which initially account for 88.4\% of the total volume, so that they take up 80\%, 60\%, 40\%, and 20\% of the total volume instead. Figure~\ref{fig:spatial} shows that \sysname consistently fulfills the highest demand across all spatial distributions. Pinning's performance decreases by 8.3\%, as its strategy of focusing only on top demands relies on the presence of a heavy-tailed demand distribution. Teal has a slight decrease of up to 2.4\% due to unseen demand distribution in training. This is a relatively smaller decrease than the temporal fluctuation, because Teal can still handle simple, evenly distributed traffic demands. POP has a slight increase of 1.4\% as demands are more granular, where no demand requires a significant fraction of the total resources.

\parab{Link failures.} Widespread failures across inter-datacenter links
are relatively rare in real-world operations unless triggered by physical
fiber cuts~\cite{singh2021cost}.
However, we simulate severe failure scenarios,
introducing 50, 100, and 200 simultaneous link failures.
As shown in Figure~\ref{fig:failures}, the satisfied demand declines consistently
across all methods.
Specifically, all approaches drop by 0.6--1.1\% under 50 link failures, 1.5--2.1\% under 100 link failures, and 3--4.2\% under 200 link failures. These drops stem from the fact that the failed links represent only a small portion of the total 8,558 links, allowing all approaches to recompute flow allocation given adequate time.

\subsection{Micro-benchmarks}\label{sec:eval_micro}

We conduct micro-benchmarks to evaluate the impact of \sysname's design components, with all experiments performed in maximizing total flow in traffic engineering.

\parab{Speedup when varying CPU cores.} Figure~\ref{fig:cpu_time} evaluates the speedup of \sysname, \sysnamestar, and Exact sol. with 1, 4, 16, and 64 CPU cores. \sysnamestar achieves 61.7$\times$ speedup under 64 cores, where the slight gap occurs as we cannot exactly divide the total computation time evenly across 64 cores. \sysname also achieves an almost linear speedup when the number of CPU cores grows from 1 to 16. However, this linearity diminishes with 64 cores, where \sysname only achieves 18.2$\times$ speedup. This is because utilizing all 64 cores for parallelism incurs overhead from cache contention and stragglers, as discussed in \S\ref{sec:eval_cluster_scheduling}. In contrast to the linear speedup of \sysnamestar and \sysname, the speedup of Exact sol. is sublinear and marginal, achieving only 3.4$\times$ speeds with 64 CPU cores. This is because of commercial solvers’ sequential nature, e.g., the simplex method~\cite{simplex} in linear programming requires thousands to millions of small steps to reach the optimal solution.

\parab{Convergence rate.}
Teal generates a coarse, initial solution to traffic engineering using neural networks
and fine-tunes it through ADMM.
While Teal also ``decouples'' constraints, it does not further ``decompose''
the problem as \sysname does.
Instead, it analytically and approximately minimizes a problem-specific
augmented Lagrangian, directly projecting the result onto the non-negative
feasible region.
As a specialized method, Teal converges rapidly,
achieving 90\% satisfied demand within 10 seconds,
as shown in Figure~\ref{fig:iter}.
In comparison, \sysname requires $\sim$30 seconds to attain comparable performance.
In each optimization interval, \sysname is warm-started using the previous solution
by default. When initialized instead
with Teal's neural network output,
\sysname converges at a similar rate.
In contrast, employing a naive initialization---where each demand is equally
split across available paths---reduces \sysname's convergence speed by nearly half.

\parab{Alternative optimization methods.} Figure~\ref{fig:decouple} evaluates alternative constrained optimization techniques for maximizing total flow in traffic engineering---the penalty method~\cite{penalty} and the augmented Lagrangian method~\cite{augmentedlgr}, comparing them with the ADMM-based approach used by \sysname.
Both methods aim to solve the reformulated optimization problem in 
Equation~\ref{eq:reformulated_opt} by jointly (and naively)
optimizing variables $x$ and $z$,
but they differ in how they enforce constraints.
The penalty method introduces penalty terms into the objective function to
discourage constraint violations. It starts with a small penalty coefficient
and gradually increases it toward infinity, thereby driving the solution toward feasibility.
However, this process typically involves solving a series of increasingly
ill-conditioned problems, resulting in slow convergence. In our evaluation,
it was more than 30$\times$ slower than \sysname in achieving a satisfied demand
of over 90\%.
The augmented Lagrangian method improves upon the penalty method by
combining both penalty terms and Lagrange multipliers.
This helps enhance convergence stability and typically reduces the number of
required iterations. 
However, unlike the ADMM approach used in \sysname, it lacks the benefit of
decomposition and parallelism. As a result, despite its faster convergence
over the basic penalty method, it remains over
3$\times$ slower than \sysname in reaching a satisfied demand of over 90\%.

\vspace{-4pt}
\section{Related Work}
\label{sec:related_work}
\vspace{-6pt}

\parab{Resource allocation problems.} Multi-tenant systems often address
resource allocation challenges. Section~\ref{sec:cases} examines three prominent examples: cluster scheduling~\cite{newell2021ras,jayaram2023sia,narayanan2020heterogeneity, xiao2018gandiva,yang2023skypilot} allocates jobs to CPUs or GPUs in heterogeneous clusters, traffic engineering~\cite{hong2013achieving, abuzaid2021contracting, zhong2021arrow, singh2021cost2, krishnaswamy2022decentralized} allocates traffic demands to network link capacities in wide-area networks, and load balancing~\cite{curino2011workload, serafini2014accordion, taft2014store} allocates query loads to servers in distributed databases.
Beyond these problems, Skyplane~\cite{jain2023skyplane} allocates cloud resources (e.g., bandwidth and TCP connections) to overlay paths that traverse intermediate cloud regions, minimizing the cost of inter-region bulk transfers.
Zeta~\cite{zhang2022zeta} allocates traffic forwarding capability from gateway clusters to virtual private clouds for scalable and robust east-west communication in large-scale clouds. Shoofly~\cite{singh2021cost} allocates light wavelengths of optical fiber to network shortcuts abstracted from optical bypasses, in order to minimize the hardware costs of provisioning long-haul capacity.
By leveraging the commonly observed separable structure of real-world resource allocation problems, \sysname is applicable to the aforementioned cases as well as a broad range of others.

\parab{Optimization algorithms.} Various constrained optimization algorithms have been employed to solve resource allocation problems, depending on the variable types, objectives, and constraints. For linear programming problems like traffic engineering, the Gurobi solver~\cite{gurobi} utilizes two primary methods. The simplex method~\cite{simplex} iteratively progresses along the boundaries of the feasible region toward the optimal solution, while the barrier method~\cite{barrier} iteratively approaches the optimal solution from within the feasible region. In mixed-integer linear programming like load balancing, the branch-and-bound method~\cite{boyd2007branch} divides the problem into smaller subproblems, or branches, and then eliminates certain branches based on bounds on the optimal solution at each iteration. For more complex optimization problems, such as cone programming when maximizing proportional fairness in cluster scheduling, the ECOS algorithm~\cite{ecos} adopts a similar iterative process as the barrier methods, with search directions found by a symmetric indefinite KKT (Karush-Kuhn-Tuck) system. All these constrained optimization algorithms operate sequentially and iteratively, and thus introduce bottlenecks in solving time. In contrast, \sysname decouples and decomposes the problem into many parallel subproblems, and thus allows faster problem solving.

\parab{Decomposition for scalability.}
As resource allocation problems grow, efforts have been made to parallelize the iterative process in optimization algorithms for better scalability.
NCFlow~\cite{abuzaid2021contracting} accelerates traffic engineering by partitioning the network into clusters and concurrently solving the flow allocation problem in each cluster. POP~\cite{narayanan2021solving} targets ``granular'' allocation problems where resources are interchangeable. It randomly splits resources and demands into subsets and allocates each subset in parallel.
Soroush~\cite{namyar2023solving} generalizes the classical waterfilling algorithm for
max-min fair allocation and presents parallelizable combinatorial algorithms with fairness guarantees.
Teal~\cite{xu2023teal} models traffic engineering with neural networks to benefit from the massive parallelism in GPU. 
Eason et al.~\cite{eason2023hose} employ the Benders decomposition algorithm to parallelize the design of a specialized cross-layer backbone network, offering performance guarantees for a particular allocation problem. While these prior decompositions rely on restrictive assumptions or problem-specific designs, \sysname offers a general and theoretically grounded approach solution for diverse resource allocation problems.

\section{Conclusion}
\vspace{-5pt}
\label{sec:conclusion}
In this paper, we scale real-world resource allocation problems
from a different lens, observing that the vast majority of them
exhibit an inherent separable structure.
This key insight allows us to systematically decouple resource
and demand constraints and decompose the original optimization problem
into independent and parallel subproblems for allocating
individual resources and demands.
We implement this approach, \sysname, as a Python package,
offering a push-button solution that accelerates large-scale resource allocation.
Through extensive evaluation across diverse resource allocation tasks,
we demonstrate that \sysname significantly improves scalability
while closely approximating the quality of exact solutions.

\vspace{-5pt}
\section*{Acknowledgments}
\label{sec:acknowledge}
We thank our anonymous reviewers and shepherd for their insightful comments.
We also thank Yang Zhou, Justin Chiu, Alexander Rush, and Victor Bahl for their valuable
feedback and support.
We extend special thanks to Suvan Chatakondu for enhancing our code after the paper's acceptance.
This work is supported in part by ACE, one of the seven centers in JUMP 2.0, a Semiconductor Research Corporation (SRC) program sponsored by DARPA. This work is also supported by National Science Foundation grant CCF-2326605 (Expedition in Computing).

\newpage
\bibliographystyle{plain}
\bibliography{references}

\appendix

\vspace{15pt}
\noindent{\large \textbf{Appendix}}
\vspace{-5pt}

\section{Evaluation Setup of Cluster Scheduling}
\label{sec:setup_cluster_scheduling}
We detail the experimental setup for the large-scale cluster scheduling evaluation
described in \S\ref{sec:eval_cluster_scheduling}.

\parab{Computing resources.} 
We collect 456 distinct GPU/CPU resource types from three benchmark sources~\cite{ignatov2019ai, lambda, EpochNotableHardwares2024}. These resource types
differ along multiple dimensions such as vendor, interconnect, memory
configuration, deployment platform, etc.
While it is unlikely today for any single cluster to host all 456 types,
recent trends in distributed ML workloads~\cite{yang2023skypilot, wu2024can}
suggest the possibility of scheduling ML jobs across global-scale clusters
in the future.
For each resource type, the number of available instances is randomly drawn
from the set $\{8, 16, 24, \ldots, 64\}$, reflecting common modern hardware configurations where nodes are provisioned with GPU/CPUs in multiples of eight.
This setup results in a total of 16,520 GPU/CPU instances in our simulated environment.

\parab{Jobs.} We generate 2,588 types of ML jobs based on the Notable AI Models dataset~\cite{EpochNotableModels2024}. For each ML model,
we synthesize multiple job types encompassing both training and inference tasks
under various numerical precisions.
The number of compute instances requested by each job is sampled from $\{1, 2, 4, 8, 16, 32\}$, in line with common practices in data and model parallelism.
Job throughput values are either taken directly from ML benchmarks~\cite{ignatov2019ai, lambda, EpochNotableHardwares2024}
or, when unavailable, estimated based on each job's FLOP requirements and
the computational capacity of the respective hardware.

\parab{Scheduling Simulator.} 
Our evaluation employs Gavel's ML scheduling simulator~\cite{narayanan2020heterogeneity}, with several adjustments.
Jobs are time-sliced across available resource types but are not co-located,
due to the increased size and resource demands of modern ML workloads.
To ensure that jobs do not grow beyond cluster capacity, we set
the average job inter-arrival time to 100 seconds in the Poisson process.
We begin the simulation from a steady state, maintaining approximately
1,500 to 2,000 active jobs at any given time.
Job allocation decisions are made every 6 minutes,
consistent with Gavel's default settings,
and the simulation runs for 200 scheduling rounds, or 20 hours.

\end{document}